\documentclass[a4paper,aps,prb, superscriptaddress,amsmath,amssymb,preprint]{revtex4-1}

\usepackage[utf8x]{inputenc}
\usepackage[english]{babel}
\usepackage{amsmath}
\usepackage{amssymb}
\newcommand{\updownarrows}{\mathbin\uparrow\hspace{-.5em}\downarrow}
\usepackage{amstext}
\usepackage{graphicx}
\usepackage{lmodern}
\usepackage{float}
\usepackage{siunitx}
\usepackage[nooneline,tight]{subfigure}
\usepackage[section]{placeins}

\usepackage[colorlinks=true]{hyperref}

\begin{document}
	
\title{Magnon mode selective spin transport in compensated ferrimagnets}
	
\author{Joel Cramer}
\affiliation{Institut f\"ur Physik, Johannes Gutenberg-Universit\"at Mainz, 55099 Mainz, Germany}
\affiliation{Graduate School of Excellence Materials Science in Mainz, 55128 Mainz, Germany}

\author{Er-Jia Guo}
\affiliation{Institut f\"ur Physik, Johannes Gutenberg-Universit\"at Mainz, 55099 Mainz, Germany}
\affiliation{Quantum Condensed Matter Division, Oak Ridge National Laboratory, Oak Ridge, 37830 TN, United States}

\author{Stephan Gepr\"ags}
\affiliation{Walther-Meißner-Institut, Bayerische Akademie der Wissenschaften, 85748 Garching, Germany}

\author{Andreas Kehlberger}
\affiliation{Institut f\"ur Physik, Johannes Gutenberg-Universit\"at Mainz, 55099 Mainz, Germany}

\author{Yurii P. Ivanov}
\affiliation{Computer, Electrical and Mathematical Sciences and Engineering Division (CEMSE), King Abdullah University 
	of Science and Technology (KAUST), Thuwal, 23955, Saudi Arabia}
\affiliation{now at: Erich Schmid Institute of Materials Science, Austrian Academy of Sciences, A-8700 Leoben, Austria}

\author{Kathrin Ganzhorn}
\affiliation{Walther-Meißner-Institut, Bayerische Akademie der Wissenschaften, 85748 Garching, Germany}
\affiliation{Physik-Department, Technische Universit\"at M\"unchen, 85748 Garching, Germany}

\author{Francesco Della Coletta}
\affiliation{Walther-Meißner-Institut, Bayerische Akademie der Wissenschaften, 85748 Garching, Germany}

\author{Matthias Althammer}
\affiliation{Walther-Meißner-Institut, Bayerische Akademie der Wissenschaften, 85748 Garching, Germany}
\affiliation{Physik-Department, Technische Universit\"at M\"unchen, 85748 Garching, Germany}

\author{Hans Huebl}
\affiliation{Walther-Meißner-Institut, Bayerische Akademie der Wissenschaften, 85748 Garching, Germany}
\affiliation{Physik-Department, Technische Universit\"at M\"unchen, 85748 Garching, Germany}
\affiliation{Nanosystems Initiative Munich (NIM), Schellingstraße 4, 80799 München, Germany}

\author{Rudolf Gross}
\affiliation{Walther-Meißner-Institut, Bayerische Akademie der Wissenschaften, 85748 Garching, Germany}
\affiliation{Physik-Department, Technische Universit\"at M\"unchen, 85748 Garching, Germany}
\affiliation{Nanosystems Initiative Munich (NIM), Schellingstraße 4, 80799 München, Germany}

\author{J\"urgen Kosel}
\affiliation{Computer, Electrical and Mathematical Sciences and Engineering Division (CEMSE), King Abdullah University 
	of Science and Technology (KAUST), Thuwal, 23955, Saudi Arabia}

\author{Mathias Kl\"aui}
\email{Klaeui@uni-mainz.de}
\affiliation{Institut f\"ur Physik, Johannes Gutenberg-Universit\"at Mainz, 55099 Mainz, Germany}
\affiliation{Graduate School of Excellence Materials Science in Mainz, 55128 Mainz, Germany}

\author{Sebastian T. B. Goennenwein}
\email{Sebastian.Goennenwein@tu-dresden.de}
\affiliation{Walther-Meißner-Institut, Bayerische Akademie der Wissenschaften, 85748 Garching, Germany}
\affiliation{Physik-Department, Technische Universit\"at M\"unchen, 85748 Garching, Germany}
\affiliation{Nanosystems Initiative Munich (NIM), Schellingstraße 4, 80799 München, Germany}
\affiliation{Institut für Festkörperphysik, Technische Universität Dresden, 01062 Dresden, Germany}

	
	\begin{abstract}
	We investigate the generation of magnonic thermal spin currents and their mode selective spin transport across interfaces in insulating, compensated ferrimagnet/normal metal bilayer systems.
	The spin Seebeck effect signal exhibits a non-monotonic temperature dependence with two sign changes of the detected voltage signals.
	Using different ferrimagnetic garnets, we demonstrate the universality of the observed complex temperature dependence of the spin Seebeck effect.
	To understand its origin, we systematically vary the interface between the ferrimagnetic garnet and the metallic layer, and by using different metal layers we establish that interface effects play a dominating role.
	They do not only modify the magnitude of the spin Seebeck effect signal but in particular also alter its temperature dependence.
	By varying the temperature, we can select the dominating magnon mode and we analyze our results to reveal the mode selective interface transmission probabilities for different magnon modes and interfaces.
	The comparison of selected systems reveals semi-quantitative details of the interfacial coupling depending on the materials involved, supported by the obtained field dependence of the signal.
	\end{abstract}

\maketitle

In recent years, the generation and flow of spin currents and thus the transport of spin information has been studied intensively as it is an essential prerequisite for the realization of spintronic devices. 
In spintronics not only the electron charge but also its spin state is used for information transmission and manipulation  \cite{Wolf2001,Zutic2004}, which promises reduced power consumption compared to conventional charge-based electronics that entails ohmic losses.
The viability of this concept has recently been corroborated for instance by the demonstration of multi-terminal logic majority gates based on the superposition of spin currents  \cite{Kostylev2005,Klingler2014,Klingler2015,Ganzhorn2016}. 

In general, spin currents can be generated in magnetic materials in different ways as, for instance, ferromagnetic resonance (FMR) based spin pumping  \cite{Tserkovnyak2002,Mizukami2002,Saitoh2006}, electrical injection by the spin Hall effect (SHE) \cite{Cornelissen2015,Goennenwein2015} or the excitation of thermal spin currents via the spin Seebeck effect (SSE)  \cite{Uchida2010,Bauer2012,Kehlberger2015}.
The induced spin currents are then detected in an adjacent normal non-magnetic metal (NM) layer by the inverse spin Hall effect (ISHE) \cite{Sinova2015}.
SSE measurements on Y\textsubscript{3}Fe\textsubscript{5}O\textsubscript{12}/Pt (YIG/Pt) bilayers with different YIG thicknesses revealed a strong thickness dependence of the effect, confirming that in ferromagnetic insulators (FMI) thermally generated spin currents are carried by magnonic excitations of the spin system \cite{Kehlberger2015}.
Beyond the key influence of bulk properties for the spin current generation and transport in insulating ferrimagnets, indications were found that the SSE signal also depends on the YIG/NM interface \cite{Aqeel2014,Qiu2015,Guo2016}.
Guo \textit{et al.}  \cite{Guo2016} have demonstrated that in different YIG/NM hybrids both the SSE amplitude and its temperature dependence are influenced by the NM material.
This suggests a more complex behavior of the SSE signal strength and sign with a possible explanation being based on different frequency and momentum dependencies of spin transmission probabilities across different interfaces to different NMs.
While different materials are shown to yield different SSE characteristics, it is not clear to what extent material properties or the morphology of the interface dominate spin transmission properties.
Previous findings have not allowed for a quantitative understanding, as the magnetic structure of YIG prohibits a distinction and separation between the different magnon modes that make up the total spin current.
Furthermore, given indications by spin pumping measurements that the signal amplitude measured only weakly depends on the excitation frequency \cite{Harii2011}, it is perceived as common wisdom that the spin transmissivity across interfaces is magnon mode independent.
The possible mechanism of different transmission amplitudes across the interface is therefore difficult to resolve as there is no means to distinguish between the modes excited.

Recent SSE experiments performed using the compensated ferrimagnet Gd\textsubscript{3}Fe\textsubscript{5}O\textsubscript{12} (GdIG), which comprises a more complex magnetic sublattice configuration, have revealed a more complicated temperature dependence of the SSE signals \cite{Geprags2016}.
When measured as a function of temperature, two sign changes of the signal amplitude appear.
The sign change occurring near the so-called magnetic compensation point, where the magnetizations of the sublattices cancel each other out, is explained by the reorientation of the magnetic sublattices \cite{Geprags2016}.
A second sign change at lower temperatures, however, is the result of a competition between two bulk magnon modes with opposite precession yielding spin currents of opposite spin polarity.
The first mode is the fundamental ferrimagnetic magnon mode exhibiting a gapless dispersion with a narrow bandwidth, while the second mode is gapped by the highly temperature dependent exchange coupling between the Gd and the Fe sublattice.
This demonstrates that the SSE is a probe of magnon properties of the respective material. While established for GdIG, it is unclear if this complex behavior is a universal phenomenon that occurs for other ferrimagnetic materials as well.
Given this particular temperature dependence, the  SSE experiments in different compensated ferrimagnets provide the unique opportunity to infer the importance of bulk vs interface effects for the spin transport across the interface.
The evolution of the SSE signal at temperatures around the low temperature SSE sign change allows one to select the dominating magnon mode and thus to probe the mode-selective spin transmission across the interface to the NM, which has not been possible so far.
Furthermore, intentional modifications of the interface as well as using different NM layers for the spin current detection yield access to the magnon selective interface-related spin transmissivity properties.

In this Letter we compare the SSE response of different rare-earth iron garnets in order to extract the mode- and interface-dependent spin transmission.
We start by checking whether the complex SSE behavior found in Gd based iron garnets indeed is a universal phenomenon for rare-earth iron garnets with varying thickness.
To isolate the interface effect, we systematically vary the interfaces of the FMI/NM bilayers by either changing the NM detection layer or using different surface treatments of the FMI before the NM deposition.
Finally, we combine this with the high magnetic field influence on the temperature dependent spin transmission efficiency and identify the spin current contribution of distinct magnon modes.

For our investigation of the temperature dependent excitation of thermal magnons we have selected ferrimagnetic garnet thin films of the stoichiometry Gd\textsubscript{3}Fe\textsubscript{5}O\textsubscript{12}, Dy\textsubscript{3}Fe\textsubscript{5}O\textsubscript{12} and (Gd\textsubscript{2}Y)(Fe\textsubscript{4}In)O\textsubscript{12} (GdYInIG).
We chose GdYInIG to study the effect of combining magnetic (Gd) and non-magnetic (YIn) rare earth ions while Dy\textsubscript{3}Fe\textsubscript{5}O\textsubscript{12} is used to investigate the effect of other magnetic rare-earth ions (Dy) compared to the previously used Gd (in contrast to Gd\textsuperscript{3+} ions, Dy\textsuperscript{3+} ions feature a finite orbital moment ($L=5$)).
All thin (\SI{20}{\nano\meter} to \SI{200}{\nano\meter}) films are grown epitaxially on either Gd\textsubscript{3}Ga\textsubscript{5}O\textsubscript{12} (GGG) or Y\textsubscript{3}Al\textsubscript{5}O\textsubscript{12} (YAG) substrates by pulsed laser deposition in an oxygen atmosphere. Afterwards, the Pt or Pd detection layer is deposited by magnetron-sputtering (ex-situ) or electron beam evaporation (in-situ) on top of the garnet thin film.
Similarly to our previous work \cite{Guo2016}, high resolution x-ray diffraction as well as transmission electron microscopy have been used in order to check for the quality of the deposited films and the interfaces between them.
X-ray reflectivity and atomic force microscopy reveal smooth surfaces of the grown garnet films (see supplemental information\cite{Supplemental}).
The respective high resolution scanning TEM (HR-STEM) images and electron energy loss spectroscopy (EELS) elemental maps clearly reveal the crystalline structure of the grown garnet films as well as well-defined interfaces between epitaxial ferrimagnetic garnet thin film and the normal metal layer.
Additionally, superconducting quantum interference device (SQUID) magnetometry measurements were carried out for each sample to determine the magnetization compensation point.
For garnet films grown on YAG substrates, the temperature dependence of the magnetization M has been measured by sweeping the system temperature while applying an external magnetic field of \text{$\mu$}\textsubscript{0}H = \SI{1}{\tesla}.
For thin films on GGG substrates, however, the large paramagnetic moment of GGG prohibits this procedure. M thus has been determined via magnetic field sweeps at constant temperatures.

\begin{figure}[p]
	\includegraphics[width = 0.75 \textwidth]{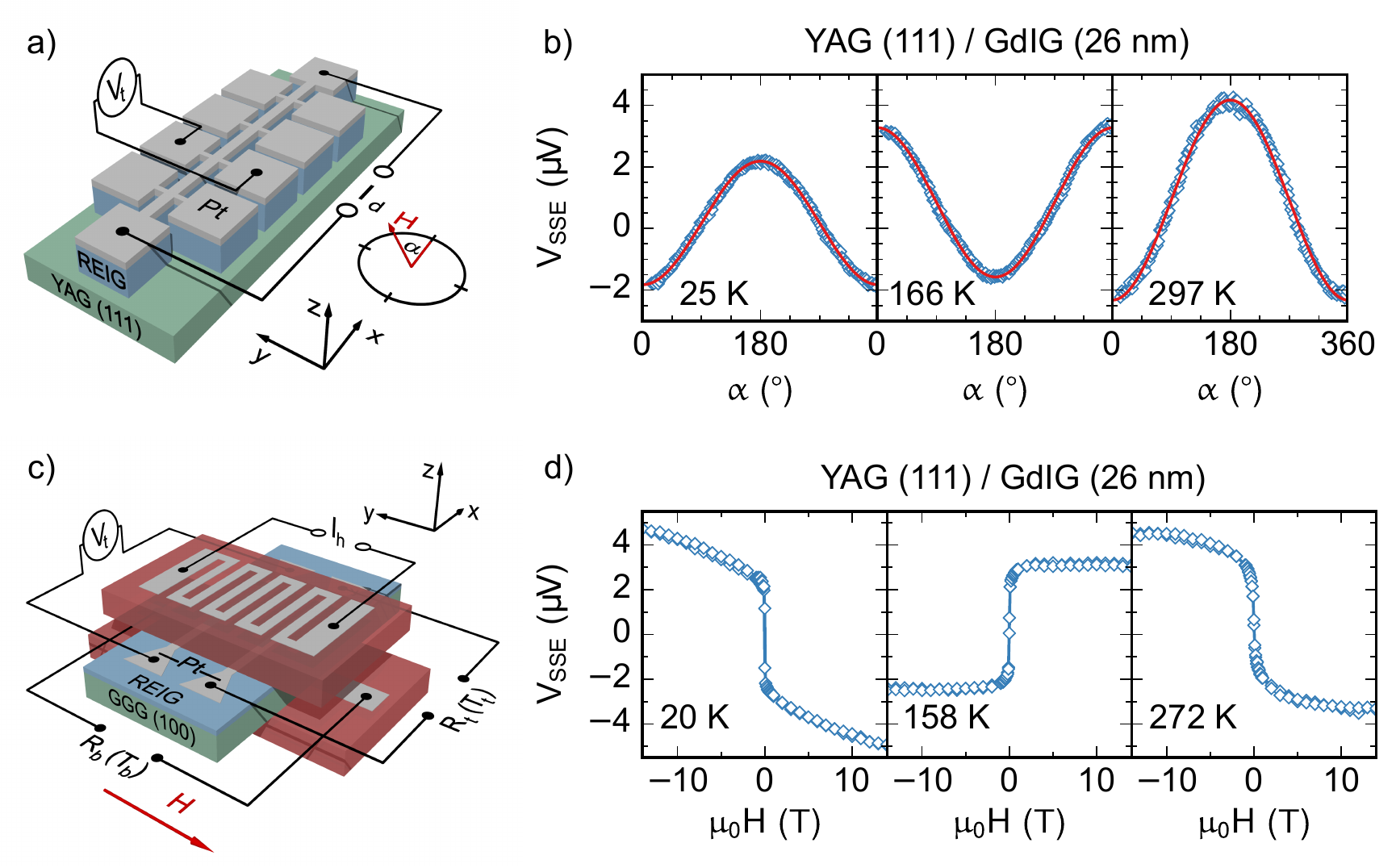}
	\caption{
		Schematics of the setups used to measure the spin Seebeck effect.
		At WMI, a) the out-of-plane temperature gradient is generated by driving a large current I\textsubscript{d} through a Hall bar mesa structure while the thermo-voltage is recorded as b) a function of the in-plane magnetic field orientation as well as d) a function of the magnetic field strength at a fixed magnetic field orientation.
		In Mainz c) the out-of-plane temperature gradient is generated by an external heater attached on top of the sample.
		The SSE amplitude is obtained by sweeping a perpendicular magnetic field.}
	\label{fig:1}
\end{figure}

SSE measurements are performed in different setup geometries and measurement methods.
Samples produced at the WMI are patterned into a Hall bar mesa structure, as shown in Fig. \ref{fig:1} a), which allows for simultaneous heating and signal detection \cite{Geprags2016,Schreier2013}.
To detect the SSE signal, either measurements as a function of the magnetic field magnitude at a fixed in-plane field orientation (hysteresis measurements) or in-plane angular-dependent measurements at a fixed magnetic field strength are performed.
The latter enable to quantify the SSE signal by the characteristic sinusoidal angular dependence, while in the hysteresis measurements the signal is extracted from the difference between positive and negative magnetic field, as shown in Fig. \ref{fig:1} b) and d) for different temperatures.
A second type of SSE setup has been used for samples fabricated and measured in Mainz.
An additional heater is placed on top of the measurement-stripe structure electrically isolated from the detection layer \cite{Geprags2016,Guo2016a}, as shown in Fig. \ref{fig:1} c).
For this setup, hysteresis measurements  \cite{Kehlberger2015} and temperature sweep measurements \cite{Guo2016a} are performed.
In the latter the SSE signal is obtained by the difference of temperature dependent measurements at a fixed positive and negative magnetic field.
All measurement methods yield a quantitatively identical SSE signal and show the characteristic linear dependence of the signal strength on the applied heating power.
Simultaneous measurements of the detection (metal) layer and bottom layer resistance ($R_t$ and $R_b$) are used to determine the temperature gradient applied to the sample stack and the actual temperature of the sample surface.
This temperature is used in the following as the reference temperature for the sample.
Note that we show here the measured spin Seebeck voltage and not the scaled data in \si{\micro\volt\per\kelvin}.
This is to obtain robust information as the flattening of the $R$ vs. $T$ dependence at very low temperatures complicates and partially prohibits the accurate determination of the temperature gradient.
Since we have measured all samples that have similar thermal properties in nominally identical conditions, a robust comparison of the data is possible.
In the following measurement data are thus plotted in units of \si{\micro\volt}.

\begin{figure}[p]
	\includegraphics[width = 0.75 \textwidth]{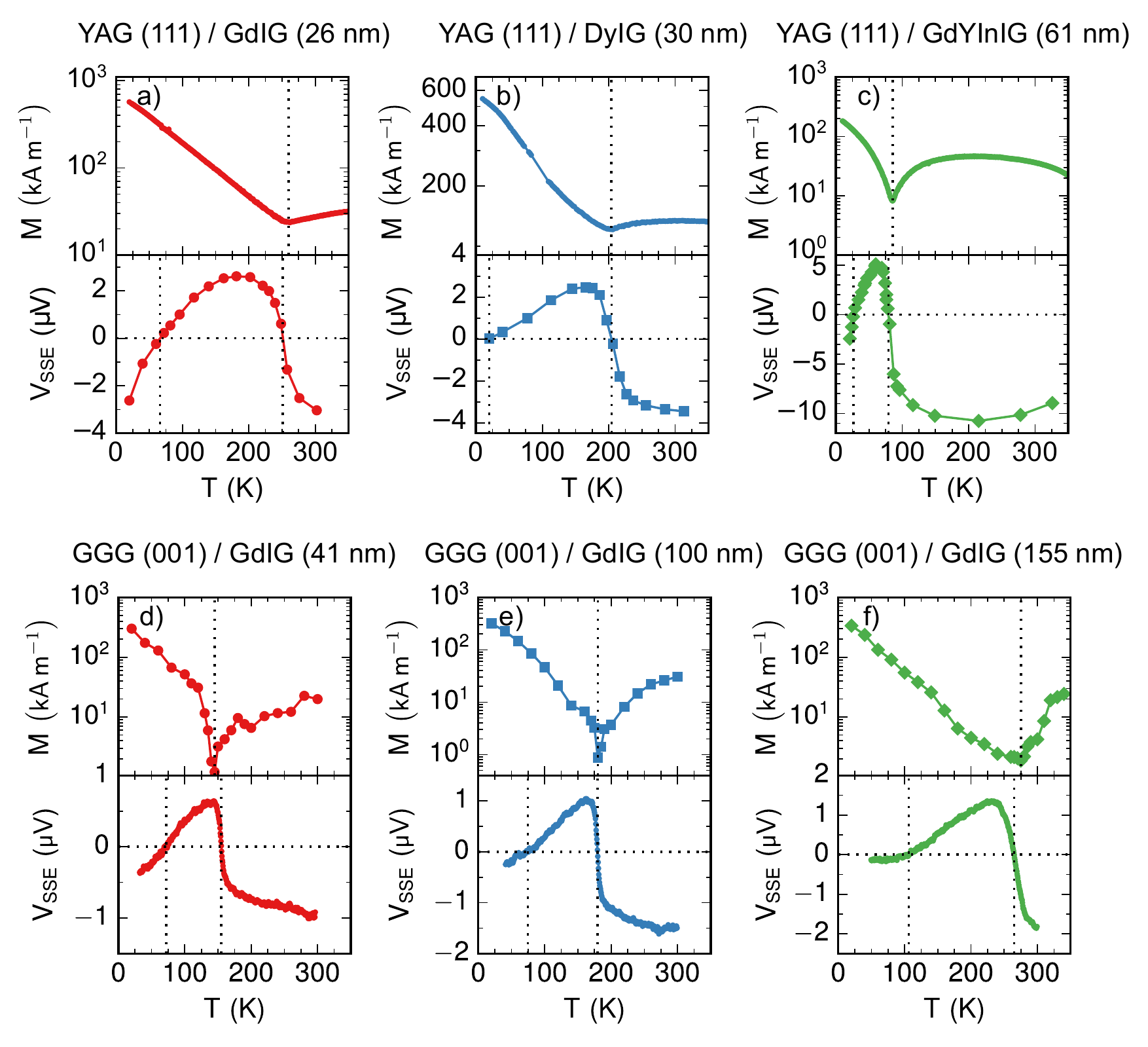}
	\caption{
		In each panel, the upper plot shows the magnetic moment of the garnet film as a function of temperature determined by SQUID magnetometry measurements.
		The SSE voltage as a function of temperature is shown in the bottom plot for the same sample. The upper three figures a), b), c) show the data of \textit{RE}IG/Pt heterostructures with \textit{RE}IG = GdIG, DyIG and GdYInIG grown on YAG.
		The lower three figures d), e) and f) contain the data for GdIG/Pt  heterostructures grown on GGG, featuring different GdIG layer thicknesses.
		The dashed lines mark the temperature of the magnetic compensation point T\textsubscript{comp} determined from SQUID magnetometry data and the temperatures at which the SSE voltage changes sign.
		For all measurements a magnetic field of \text{$\mu$}\textsubscript{0}H = \SI{1}{\tesla} is applied.}
	\label{fig:2}
\end{figure}

All SSE measurement methods consistently yield two sign changes of the SSE voltage amplitude as a function of temperature for the garnets studied here (Fig. \ref{fig:2}).
At 300 K, the signal has the same sign as the SSE voltage commonly observed in YIG/Pt \cite{Kehlberger2014,Schreier2015}.
Upon decreasing the temperature, the signal amplitude decreases, resulting in a first SSE voltage sign change.
At this first sign change the SSE-hysteresis flips orientation as shown in Fig. \ref{fig:1} d).
The temperature of the first sign change T\textsubscript{sign,1} matches with the temperature of the magnetic compensation point T\textsubscript{comp} within the error bar for all investigated garnets, as can be seen in Fig. \ref{fig:2} a)-c).

\begin{figure}[p]
	\includegraphics[width = 0.75 \textwidth]{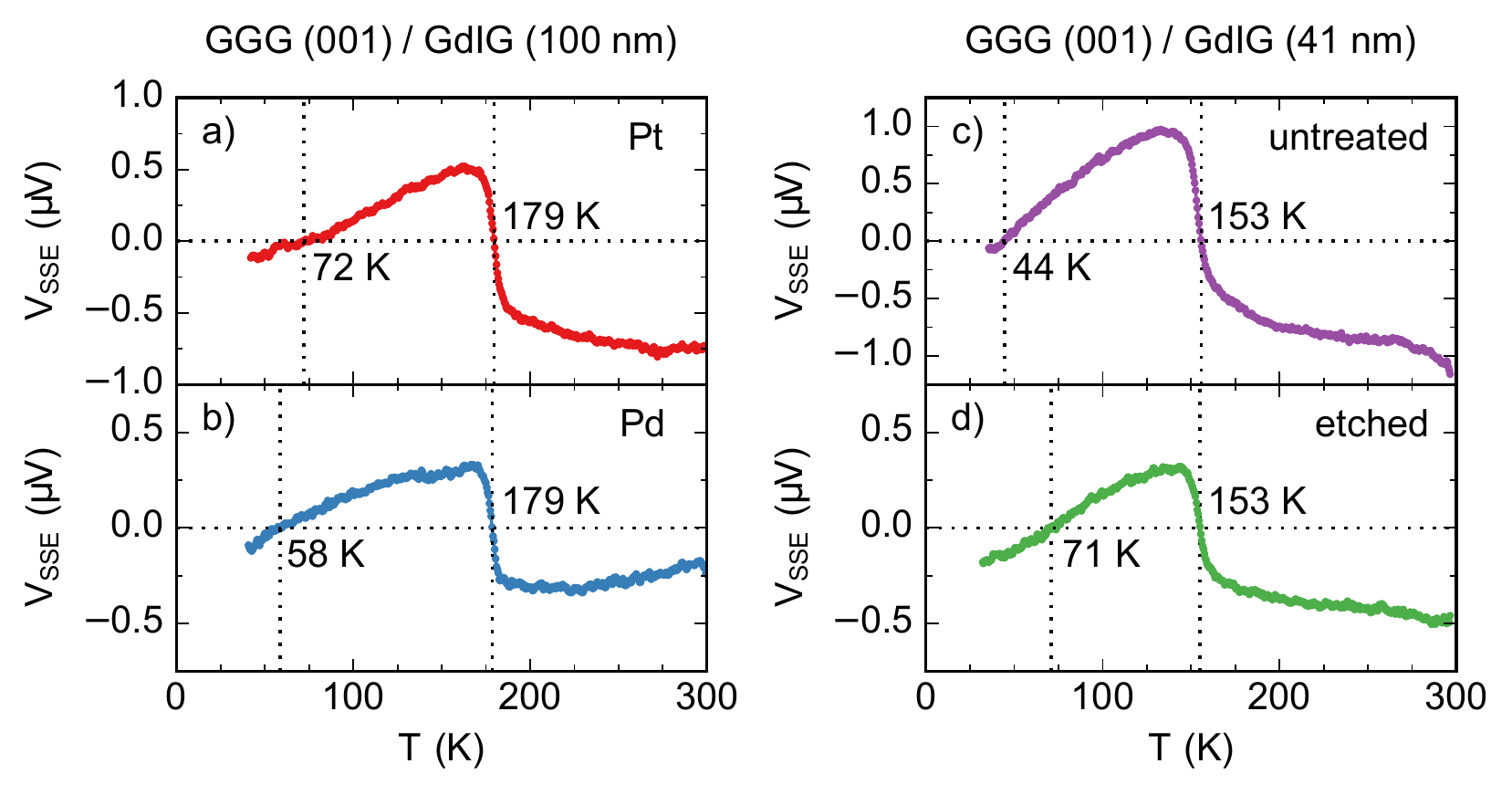}
	\caption{
		SSE voltage as function of temperature measured in GdIG/NM heterostructures grown on GGG (001).
		(a), (b) V\textsubscript{SSE} of GdIG/NM heterostructures using Pt as well as Pd as detection layer.
		The same GdIG thin film sample was used for both heterostructures.
		(c), (d) V\textsubscript{SSE} of two GdIG/Pt heterostructures.
		The Pt thin films were deposited on both an untreated and etched surface of the same GdIG thin film sample.
		The dashed lines mark the temperatures at which the SSE voltage changes sign.}
	\label{fig:3}
\end{figure}

We further investigate the GdIG thickness dependence of this sign change of the SSE voltage.
We find that the temperature for magnetic compensation, T\textsubscript{comp}, changes with the thickness of our GdIG films grown epitaxially on GGG as shown in Fig. \ref{fig:2} d)-f).
For increasing film thickness we observe a shift of the compensation point temperature towards the reported bulk value of \SI{286}{\kelvin} \cite{Geller1965}.
The shift of the compensation point T\textsubscript{comp} to higher temperatures for increasing the GdIG film thickness is reflected in the shift of the first sign change of the SSE signal as T\textsubscript{sign,1} coincides with T\textsubscript{comp} for all investigated samples. 
As the temperature is decreased further, a second sign change of the SSE voltage is observed in all investigated samples at T\textsubscript{sign,2}.
Around T\textsubscript{sign,2} a continuous increase of the net magnetic moment is observed, indicating that no qualitative change of the sublattice magnetizations takes place in this temperature range.
This is further strengthened by recent observations made by Ganzhorn \textit{et al.} \cite{Ganzhorn2016Canted}, where the spin Hall magnetoresistance (SMR) has been investigated in GdYInIG/Pt bilayers as a function of temperature.
The SMR reveals a distinct dependence on the orientation of the magnetic sublattices with the result that in the vicinity of T\textsubscript{comp}, where a canted spin state appears, the SMR changes sign.
In the region of T\textsubscript{sign,2}, however, the SMR does not exhibit unexpected behavior.
Generally, the temperature of the second sign change T\textsubscript{sign,2} increases slightly as the temperature of the first sign change T\textsubscript{sign,1} increases, however, there is no direct proportionality of T\textsubscript{sign,1} and T\textsubscript{sign,2} (cf. Fig. \ref{fig:2} d)-f)).
Two possible explanations were put forward to explain the low temperature sign change in GdIG \cite{Geprags2016}.
The first explanation is based on a competition of the two dominating bulk magnon modes with temperature and the second one on different interfacial magnon mode transmission probabilities at the GdIG/Pt interface.
To check the relevance of these two different models, we investigate the influence of different detection layer materials on the SSE signal.
For this purpose we cleave a GdIG/GGG (100) sample into two parts, which allows us to ensure identical GdIG thin film properties.
The two equal GdIG thin films are then covered by either Pt or Pd layers.
The HR-STEM analysis indicates a very good epitaxial quality of the GdIG film as well as a high surface flatness of both metallic films.
In addition to that, EELS elemental maps reveal similar sub-nanometer interfaces between Pt or Pd and GdIG.
Fig. \ref{fig:3} a) and b) show the SSE signal as a function of temperature for both samples.
While the first sign change occurs at the same temperature T\textsubscript{sign,1}, as expected for one and the same GdIG thin film if the bulk magnetic properties are decisive, the temperature of the second low temperature sign change at T\textsubscript{sign,2} depends strongly on the NM layer.
Compared to the Pt capped sample, in the sample with Pd as detection layer T\textsubscript{sign,2} is shifted to lower temperatures and the overall signal amplitude is reduced.

To probe the influence of the interface quality on the SSE signal, a second GdIG/GGG sample is cleaved into two parts, whose surfaces are treated differently before the deposition of nominally identical Pt detection layers.
One sample is only treated by acetone and isopropyl alcohol cleaning prior the deposition (marked as “untreated”), while the other sample is etched by an O-ion bombardment in-situ prior to the Pt deposition (marked as “etched”).
As shown in Figs. \ref{fig:3} c) and d), again, T\textsubscript{sign,1} remains unaltered while T\textsubscript{sign,2} changes.
Compared to the untreated sample, T\textsubscript{sign,2} of the etched sample is shifted to a higher temperature accompanied by an overall reduced signal amplitude.
These results demonstrate the strong influence of the interface quality on the temperature of the second sign change.

\begin{figure}[p]
	\includegraphics[width = 0.75 \textwidth]{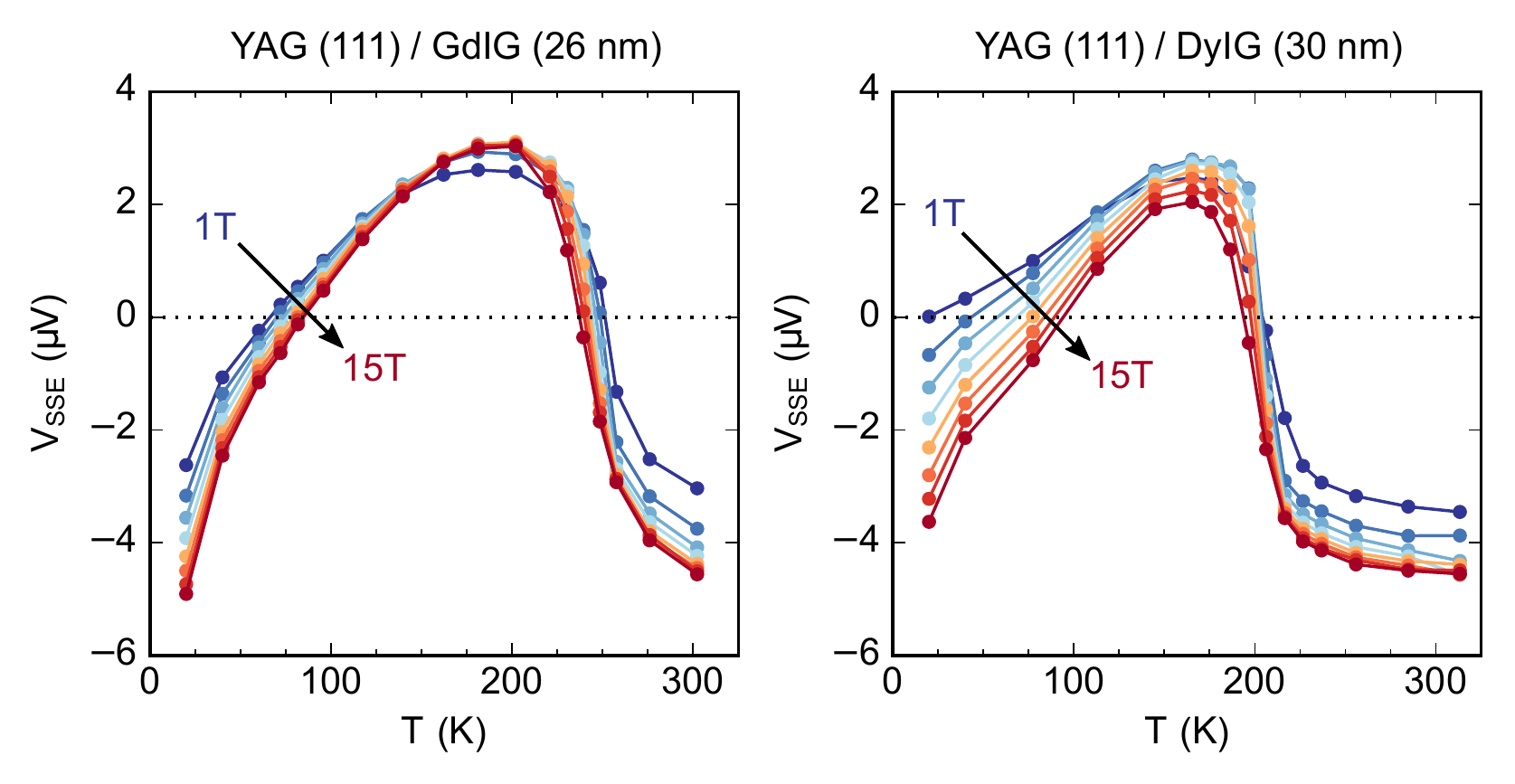}
	\caption{
		SSE voltage as function of temperature for (a) GdIG/Pt and (b) DyIG/Pt heterostructures for different magnetic field strengths.}
	\label{fig:4}
\end{figure}

As shown by Kikkawa and coworkers \cite{Kikkawa2015}, the magnetic field strength has a significant influence on the SSE amplitude.
To further investigate this matter in compensated garnet/NM heterostructures, we probe the dependence of the magnetic field on the SSE signal.
Fig. \ref{fig:4} shows the SSE voltage as a function of temperature for different magnetic field strengths in GdIG/Pt and DyIG/Pt heterostructures.
T\textsubscript{sign,1} slightly shifts towards lower temperatures as the magnetic field strength is increased.
While for GdIG the temperature shift of T\textsubscript{sign,2} with magnetic field strength is similar to the temperature change of change T\textsubscript{sign,1}, a more pronounced temperature shift of the T\textsubscript{sign,2} can be observed for DyIG.
The sign change even vanishes at low magnetic field strengths, indicating a more complex SSE in DyIG as compared to GdIG.
Above the compensation point, both GdIG and DyIG reveal an increase of the SSE signal with increasing external field.

We now analyze the experimental findings presented above, considering that the SSE voltage signal can be decomposed into a contribution arising from (i) the temperature dependent dominating magnon modes within the ferrimagnet \cite{Geprags2016} and (ii) the interfacial spin transmission across the garnet/metal interface.
In particular, the comprehensive investigation with different magnetic garnet materials, different metal layers and different interface preparation procedures enables us to determine how the observed interface and magnetic field effects reveal the influence of the interfacial boundary between ferrimagnet and metal.
As shown above, the temperature of the first sign change T\textsubscript{sign,1} coincides with the temperature of the magnetic compensation point determined by SQUID magnetometry measurements within the experimental error for all investigated garnets, independent of the metal detection layer or interface quality.
Especially the interface quality and metal detection layer insensitivity of T\textsubscript{sign,1} confirm the interpretation of the origin of this sign change put forward in Ref. \cite{Geprags2016}, based on the theoretical proposal of Ohnuma \textit{et al.} \cite{Ohnuma2013}.
All investigated garnets consist of at least three coupled magnetic sublattices. Two of these sublattices (a and d lattice site) are occupied by Fe ions, which are antiferromagnetically coupled.
This sublattice configuration is also present in the classical ferrimagnetic insulator YIG, which shows no magnetic compensation point. 
By adding a further magnetic material on the c lattice site (e.g., Gd in case of GdIG), a third magnetic sublattice is introduced.
For Gd or Dy on the c lattice site, this c sublattice couples ferromagnetically to the Fe on the a lattice site  \cite{Geprags2016,Kehlberger2015a}.
Furthermore, this third sublattice possesses a strong temperature dependent magnetic moment, leading to a temperature below the Néel point (the magnetic ordering temperature) at which the net magnetic moment of the material is zero.
The temperature T\textsubscript{comp} of this so-called magnetic compensation point depends on the magnetic ions on the c sublattice as well as on their temperature dependence \cite{Geller1965}.
In a finite magnetic field the magnetic moments invert their orientation when the temperature is swept across T\textsubscript{comp}.
As a result, the sign of the SSE voltage is inverted if measured above and below the compensation point, as the polarization of the emitted magnonic spin current follows the orientation of the magnetic lattices  \cite{Geprags2016}.
Any influence of the metal detection layer or the quality of the interface on T\textsubscript{sign,1} would be evidence that interface - and not bulk magnetic properties – are relevant for T\textsubscript{sign,1}.
The fact that we observe T\textsubscript{comp} $\approx$ T\textsubscript{sign,1} irrespective of the thicknesses of the garnet layer, its composition, and its interface properties is a strong evidence that the sign change of the SSE voltage at T\textsubscript{sign,1} indeed simply reflects the change in the ‘bulk’ sublattice moment orientations viz. bulk magnon properties.
It must be noted that besides the magnetic compensation point, compensated ferrimagnets can also exhibit an angular momentum compensation point.
The theoretical model by Ohnuma \textit{et al.} \cite{Ohnuma2013} however does not predict any change of the thermal spin current due to the latter.
Indeed, we do not observe any features in our experimental data which would point to a modification of the thermal spin currents around the angular momentum compensation point.
While in GdIG the angular momentum of Gd is quenched such that magnetic compensation and angular momentum compensation temperature coincide, there would be a distinct difference between the two in DyIG.

A more complex behavior must be invoked to understand the second SSE voltage sign change at lower temperatures (T\textsubscript{sign,2}), as we discuss next.
As mentioned above, no changes of the net magnetization are observed in the vicinity of T\textsubscript{sign,2} that could explain the observed SSE behaviour.
However, the shift of T\textsubscript{sign,2} with both the material of the detection layer (Fig. \ref{fig:3} a) and b)) and the garnet surface treatment (Fig. \ref{fig:3} c) and d)) demonstrates the influence of the interface, which was previously neglected.
Expanding the original theory invoked by Ohnuma \textit{et al.} \cite{Ohnuma2013}, T\textsubscript{sign,2} is explained by a temperature dependent competition between two bulk magnonic modes in the garnet.
While at temperatures below T\textsubscript{sign,2} the spin current is mainly given by the low-energy mode (acoustic like mode), the second mode (optical like mode) becomes dominating above T\textsubscript{sign,2}.
Since the two modes are of opposite precession, their contributions to the spin current are of opposite spin polarity which results in the sign change of the SSE signal at T\textsubscript{sign,2}.
A more detailed description of this process is provided by Gepr\"ags \textit{et al.}\cite{Geprags2016}.
The efficiency of the spin transfer from the garnet into the metal detection layer, which is described by the element specific spin mixing conductance G$_{\updownarrows}$ \cite{Weiler2013}, depends on the detection layer material.
Due to a generally lower spin mixing conductance as well as spin Hall angle of Pd as compared to Pt \cite{Tao2014}, the overall signal is reduced as seen in Fig. \ref{fig:3} c)-d).
Nevertheless, the observed shift of the second sign change to lower temperatures as compared to Pt indicates that magnons of the high energy mode can be transferred more efficiently across the interface into Pd.
This is either attributable to intrinsic properties of the metal detection layer or the chemical surface termination of the garnet film.
While in case of GdIG/Pt a Gd/Fe rich interface is observed, GdIG/Pd exhibits a Gd/O rich interface (see supplemental information\cite{Supplemental}).
One thus might speculate that the decreased Fe concentration at the GdIG/Pd interface results in an increased interface damping of the low energy mode.
Besides changing the metal detection layer the interface transparency is manipulated via O-ion bombardment, yielding an universal reduction of the signal strength.
This points to a reduction of G$_{\updownarrows}$ for both modes into the detection layer.
However, as compared to the unetched interface the second sign change is found at higher temperatures, suggesting that the high energy mode is more affected by the reduced interface transparency.

Finally we corroborate the magnon mode sensitivity of the second sign change by analyzing the magnetic field dependence shown in Fig. \ref{fig:4}.
For YIG it has been shown that high magnetic fields cause a reduction of the magnonic propagation length, which is relevant for the whole temperature range \cite{Ritzmann2015}.
An energy gap (Zeeman effect) is opened by the magnetic field, which suppresses the excitation of magnons with energies lower than the gap  \cite{Kikkawa2015,Ritzmann2015,Jin2015}.
The field gap is especially important at low temperatures, as in YIG the field induced energy gap is in the order of a few Kelvin \cite{Jin2015}.
It therefore causes a shift of the whole dispersion relation to higher energies, thus mainly introducing a constant offset with respect to the zero field state \cite{Ritzmann2015}.
The resulting shift of the dispersion relation towards higher energies moves T\textsubscript{sign,2} to higher temperatures, as it now requires more thermal energy to sufficiently occupy the high energy mode.
The different sensitivities of T\textsubscript{sign,2} on the magnetic field for GdIG and DyIG might originate from the different electronic states of the Gd\textsuperscript{3+} and Dy\textsuperscript{3+} ions.
In contrast to Gd\textsuperscript{3+}, Dy\textsuperscript{3+} has a finite orbital moment ($L=5$), which may lead to differences in the magnetic field dependences of magnon modes due to the different multiplett states occupied.
The comparison to field dependent magnetization data (see supplemental information\cite{Supplemental}) shows that the increase of $V_{\mathrm{SSE}}$ at high magnetic fields above the compensation point cannot be explained by an increase of the magnetization.
Instead, one has to consider that in contrast to YIG, where the thermal signal is governed by the fundamental ferrimagnetic magnon mode, in compensated iron garnets the spin current reflects the field dependence of a complex magnon mode structure.
The quantification of the latter as well as the magnon mode dependence on different electronic states of the magnetic ions require further theoretical input, which is beyond the scope of this work.

In conclusion, we have shown that the ferrimagnetic nature and magnon mode structure of compensated garnets can be accessed by temperature dependent SSE measurements.
Our study of different compensated iron garnets, detection metal layer materials and interface qualities allows us to clearly show the universal correlation between the magnetic compensation point and the sign change of the SSE voltage at high temperatures, T\textsubscript{sign,1}.
We find that the second sign change at the lower temperature T\textsubscript{sign,2} is strongly influenced by the detection layer material as well as the interface nature, which can be rationalized in terms of a modified coupling strength of the two magnon modes.
We study these individually by varying the temperature to identify the mode selective magnon transmissivity.
The magnetic field dependence of T\textsubscript{sign,2} is consistent with our developed magnonic mode selective picture of the SSE.
T\textsubscript{sign,2} is shifted towards higher temperatures for higher magnetic fields.
Our results reveal that the SSE provides an elegant solution to gain insight into the magnon mode structure of a ferrimagnet and, on the other hand, show for the first time evidence for an interface sensitive magnon mode selective coupling towards normal metal layers.

\section{Acknowledgments}
The authors would like to express their gratitude to Prof. Gerrit E.W. Bauer and Dr. Joe Barker for valuable discussions and the Institute for Materials Research at Tohoku University for the hospitality during a visiting professor stay (MK).
Furthermore, they would like to thank Prof. Kathrin Dörr for the help in sample preparation.
This work was supported by Deutsche Forschungsgemeinschaft (DFG) SPP 1538 “Spin Caloric Transport,” the Graduate School of Excellence Materials Science in Mainz (MAINZ), and the EU projects (IFOX, NMP3-LA-2012246102, INSPIN FP7-ICT-2013-X 612759).

\bibliography{bibliography}

\pagebreak

\pagebreak

\setcounter{equation}{0}
\setcounter{figure}{0}
\setcounter{table}{0}

\renewcommand{\thesection}{S\arabic{section}}
\renewcommand{\thesubsection}{S\arabic{section}.\arabic{subsection}}
\renewcommand\thefigure{S\arabic{figure}}
\renewcommand\thetable{S\arabic{table}}  

\FloatBarrier

\begin{center}
	\LARGE Magnon mode selective spin transport in compensated ferrimagnets \vspace{0.5cm}
	\linebreak {\Large Supplemental Information}
\end{center}

\section{Sample characterization}

\subsection{X-ray diffraction \& Reciprocal space mapping}

\begin{figure}[!t]
	\centering
	\includegraphics[width = 0.8\textwidth]{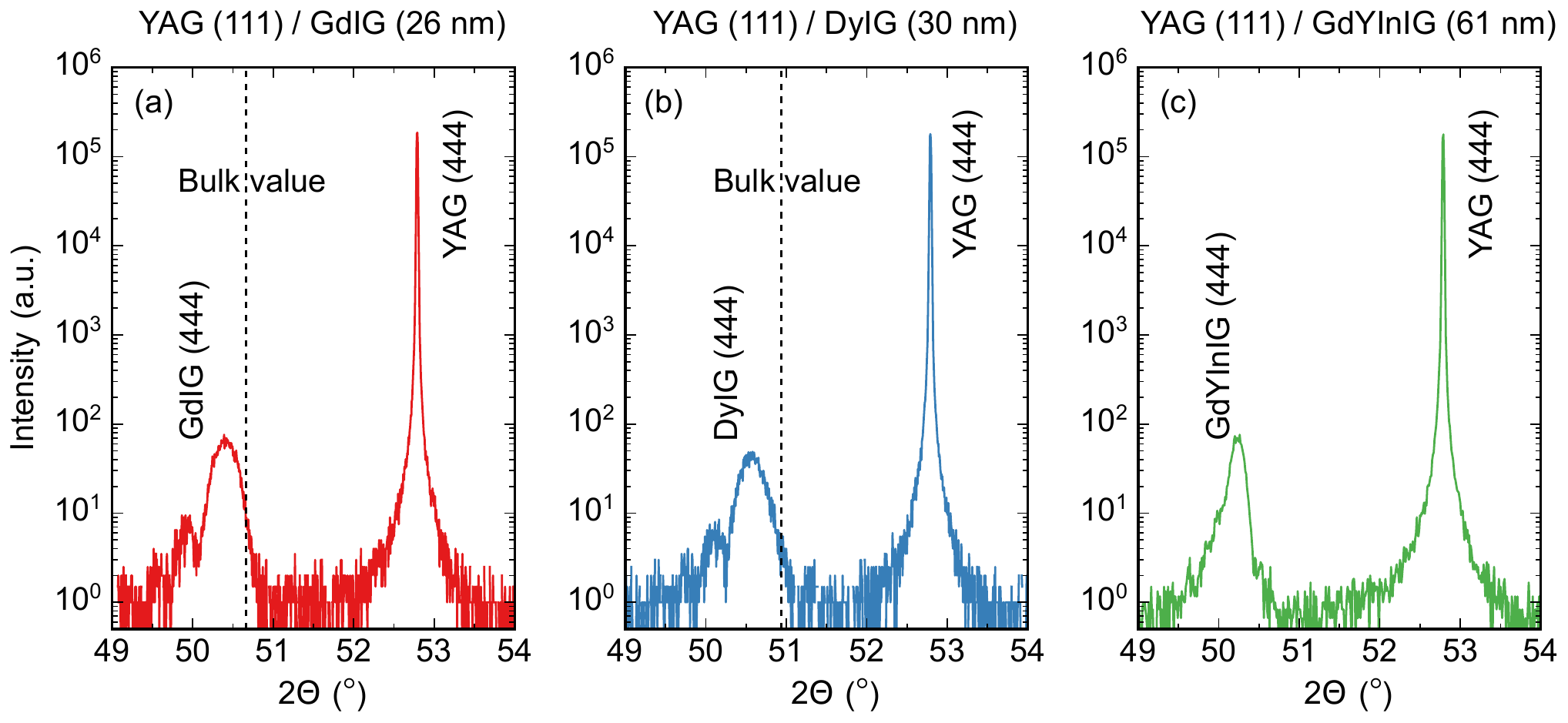}
	\caption{
		X-Ray diffraction data of the rare earth iron garnets Gd\textsubscript{3}Fe\textsubscript{5}O\textsubscript{12} (GdIG), Dy\textsubscript{3}Fe\textsubscript{5}O\textsubscript{12} (DyIG), and (Gd\textsubscript{2}Y\textsubscript{1})(Fe\textsubscript{4}In\textsubscript{1})O\textsubscript{12} (GdYInIG).
		The films are grown epitaxially via pulsed laser deposition (PLD) in an oxygen atmosphere on Y\textsubscript{3}Al\textsubscript{5}O\textsubscript{12} (YAG) substrates with (111) surface orientation.
		The vertical dashed lines indicate the (444) reflection calculated from the respective bulk lattice constant.}
	\label{fig:xrd_stephan}
\end{figure}

The crystalline quality of the investigated rare earth iron garnet films has been examined via X-ray diffractometry (XRD) and reciprocal space mapping (RSM).
In Fig. \ref{fig:xrd_stephan}, $2\Theta - \omega$ scans of the garnet films grown at the WMI are depicted, each showing the YAG (444) reflection as well as the respective \textit{RE}IG (444) reflection (\textit{RE}IG = GdIG, DyIG, and GdYInIG).
The presence of the latter indicates crystalline growth of the films, reinforced by the appearance of Laue fringes for GdIG and DyIG signifying high crystalline quality.
For GdIG and DyIG the (444) reflection appears at a scattering angle which is below the theoretical value obtained for the lattice constants $a_{\mathrm{GdIG}} = \SI{12.471}{\angstrom}$, $a_{\mathrm{DyIG}} = \SI{12.405}{\angstrom}$ \cite{Geller1969,Espinosa1962} and the Bragg equation
\begin{align*}
2d_{hkl} \sin \left( \Theta \right) &= \lambda\\
\mathrm{with}\: d_{hkl} &= \frac{a}{\sqrt{h^2 + k^2 + l^2}},
\end{align*}

assuming a cubic geometry.
The shift to lower scattering angles signifies an enhancement of the out-of-plane lattice constant, which in turn can be explained by an in-plane compressive strain considering the lattice constant $a_{\mathrm{YAG}} = \SI{12.008}{\angstrom}$ of the substrate \cite{Geller1969}.
To the knowledge of the authors there are no reported values of $a$ for GdYInIG, but regarding its close chemical relation to GdIG a similar lattice constant and thus a strained state is expected.

\begin{figure}[!t]
	\centering
	\includegraphics[width = 0.8\textwidth]{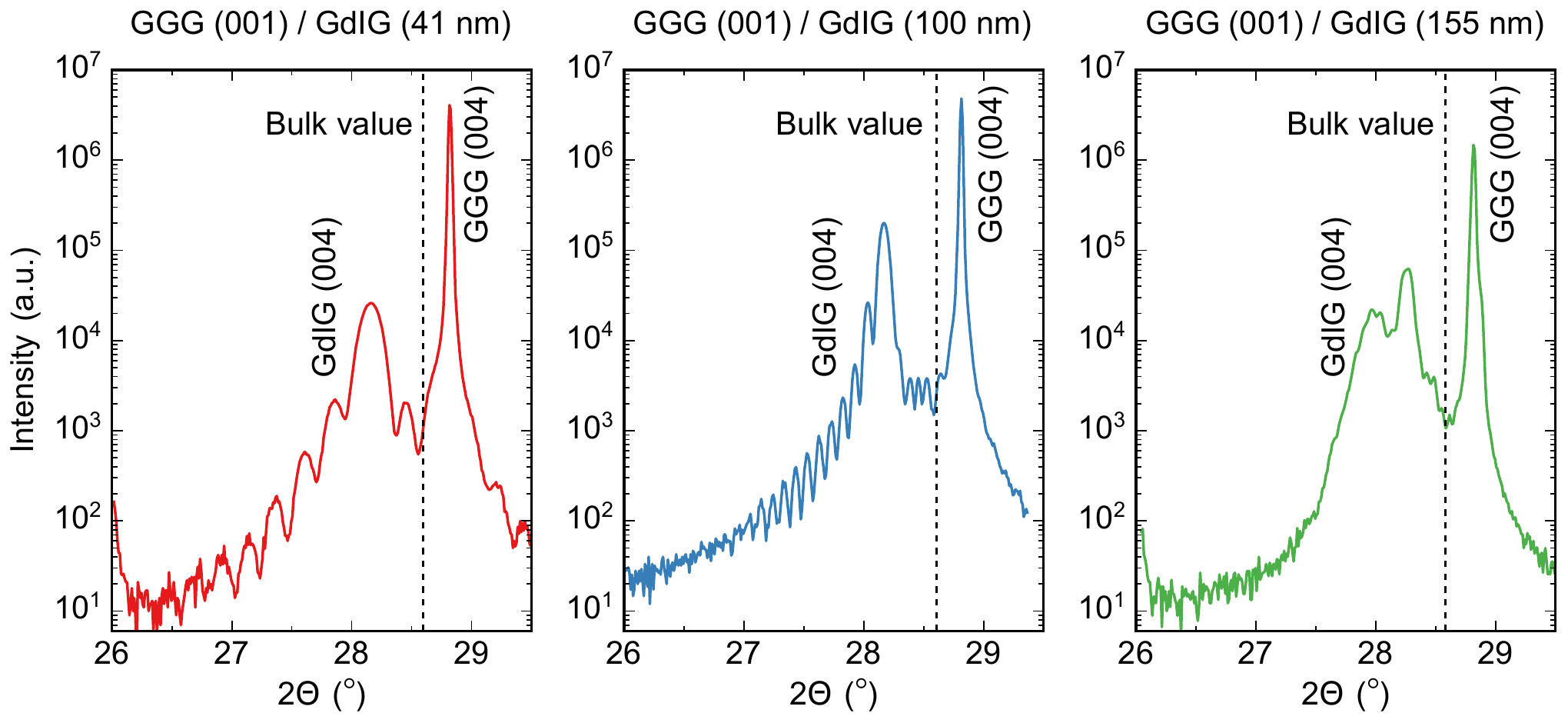}
	\caption{
		X-Ray diffraction data of Gd\textsubscript{3}Fe\textsubscript{5}O\textsubscript{12} (GdIG) thin films of various thickness (\textit{d} = \SIlist{41;100;155}{\nano\meter}). The films are grown epitaxially via pulsed laser deposition (PLD) in an oxygen atmosphere on Gd\textsubscript{3}Ga\textsubscript{5}O\textsubscript{12} (GGG) susbtrates with (001) surface orientation.
		The vertical dashed lines indicate the GdIG (004) reflection calculated from the respective bulk lattice constant.}
	\label{fig:xrd_erjia}
\end{figure}

\begin{figure}[!h]
	\centering
	\subfigure[GGG/GdIG(\SI{41}{\nano\meter})]{\includegraphics[width=0.3\textwidth]{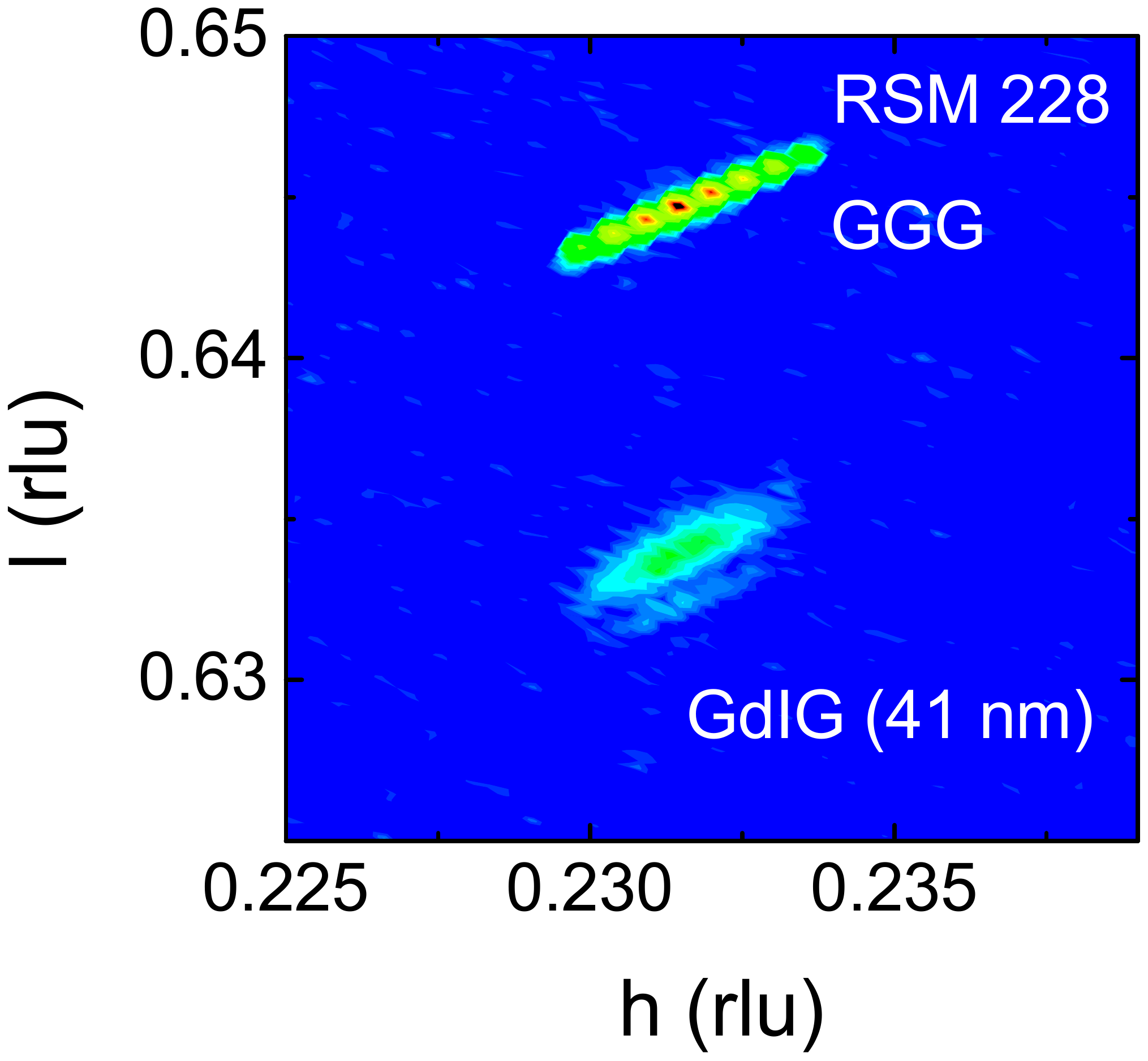}}\hspace{1cm}
	\subfigure[ GGG/GdIG(\SI{155}{\nano\meter})]{\includegraphics[width=0.3\textwidth]{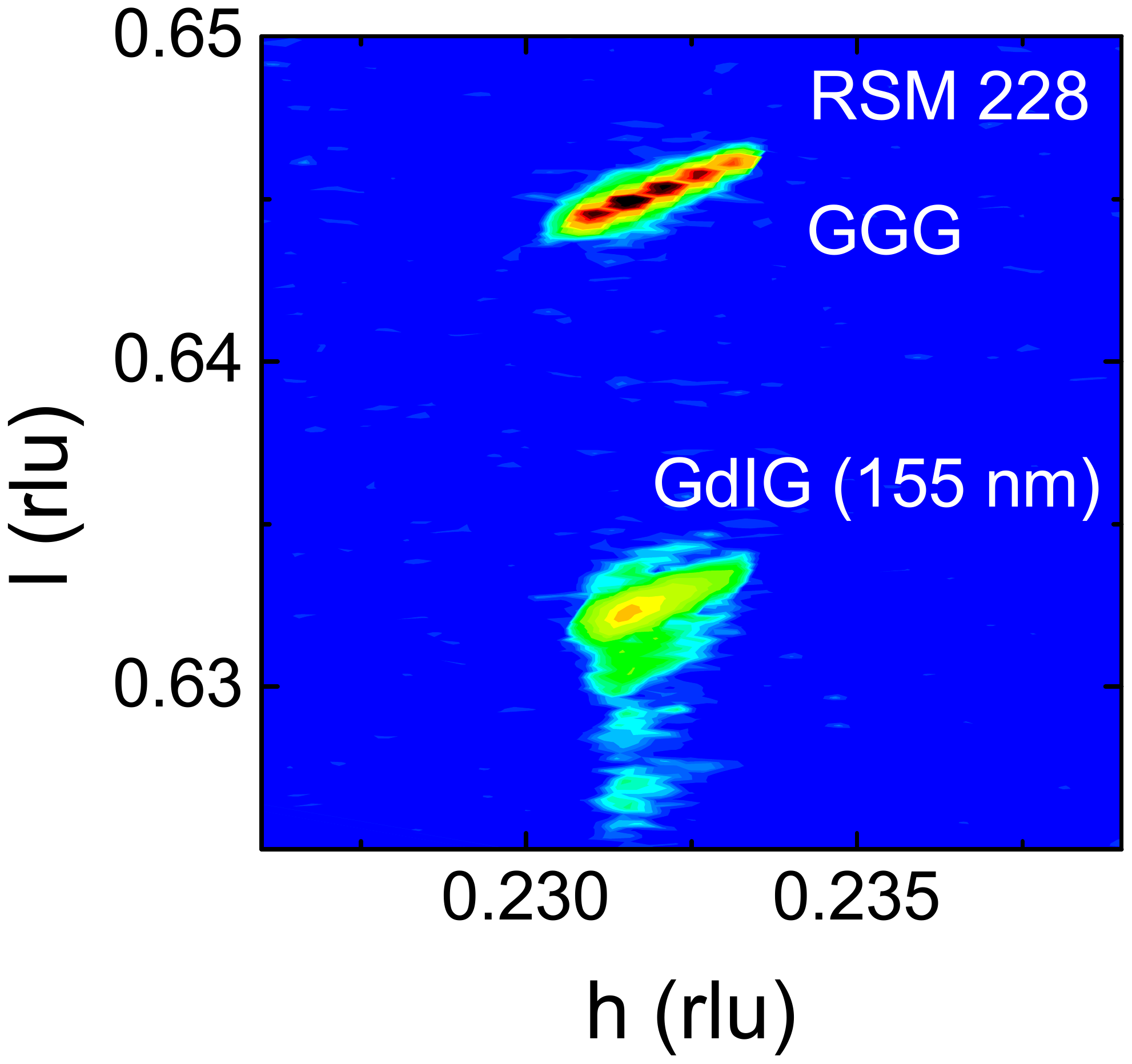}}
	\caption{Reciprocal space maps (RSM) around the GGG (228) reflection for GGG/GdIG(\SI{41}{\nano\meter}) and GGG/GdIG(\SI{155}{\nano\meter}).}
	\label{fig:rsm_gdig}
\end{figure}

XRD data of the GdIG films investigated at JGU Mainz are shown in Fig. \ref{fig:xrd_erjia}.
Since in this case the GdIG films of varying thickness are grown on GGG susbtrates with (001) orientation, the $2\Theta - \omega$ scans cover the GGG (004) and GdIG (004) reflection.
Again, the presence of the reflection peaks and the clear appearance of Laue fringes for GdIG with thickness \textit{d} = \SIlist{41;100}{\nano\meter}) indicates high crystalline quality of the films.
Furthermore, as in the case of YAG, the lattice constant of GGG $a_{\mathrm{GGG}} = \SI{12.375}{\angstrom}$ is smaller than the bulk value of $a$ for GdIG, leading to in-plane compressive stress and a shift of the scattering angle of GdIG to lower values.
The coherent growth of high quality GdIG films up to \SI{155}{\nano\meter} is further emphasized by the reciprocal space maps around the GGG (228) reflection presented in Fig. \ref{fig:rsm_gdig}.

\pagebreak
\FloatBarrier

\subsection{X-ray reflectometry \& Atomic force microscopy}

\begin{figure}[bt]
	\centering
	\includegraphics[width = 0.8\textwidth]{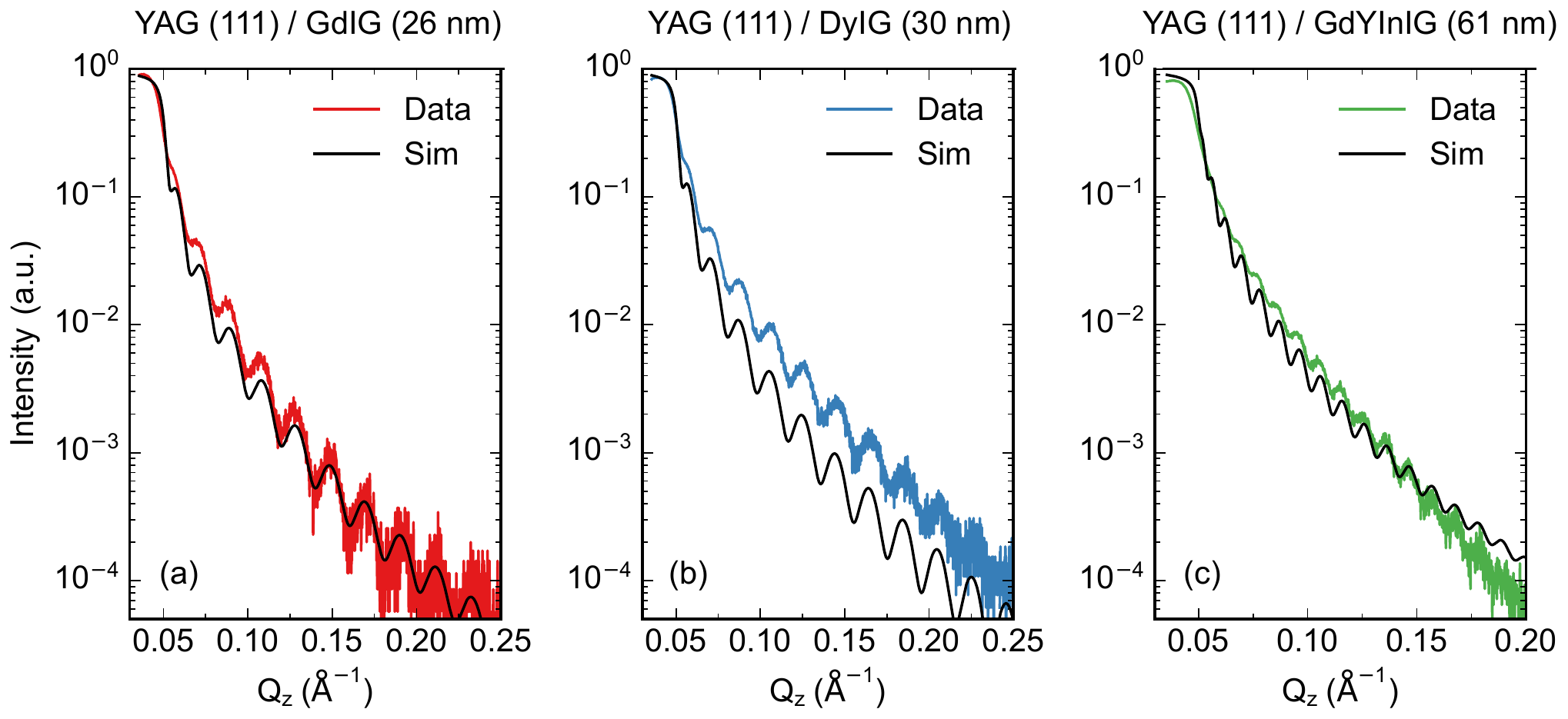}
	\caption{
		X-Ray reflectometry data of rare earth iron garnets Gd\textsubscript{3}Fe\textsubscript{5}O\textsubscript{12} (GdIG), Dy\textsubscript{3}Fe\textsubscript{5}O\textsubscript{12} (DyIG), and (Gd\textsubscript{2}Y\textsubscript{1})(Fe\textsubscript{4}In\textsubscript{1})O\textsubscript{12} (GdYInIG) investigated at the WMI.}
	\label{fig:xrr_stephan}
\end{figure}

\begin{figure}[!bt]
	\centering
	\includegraphics[width = 0.8\textwidth]{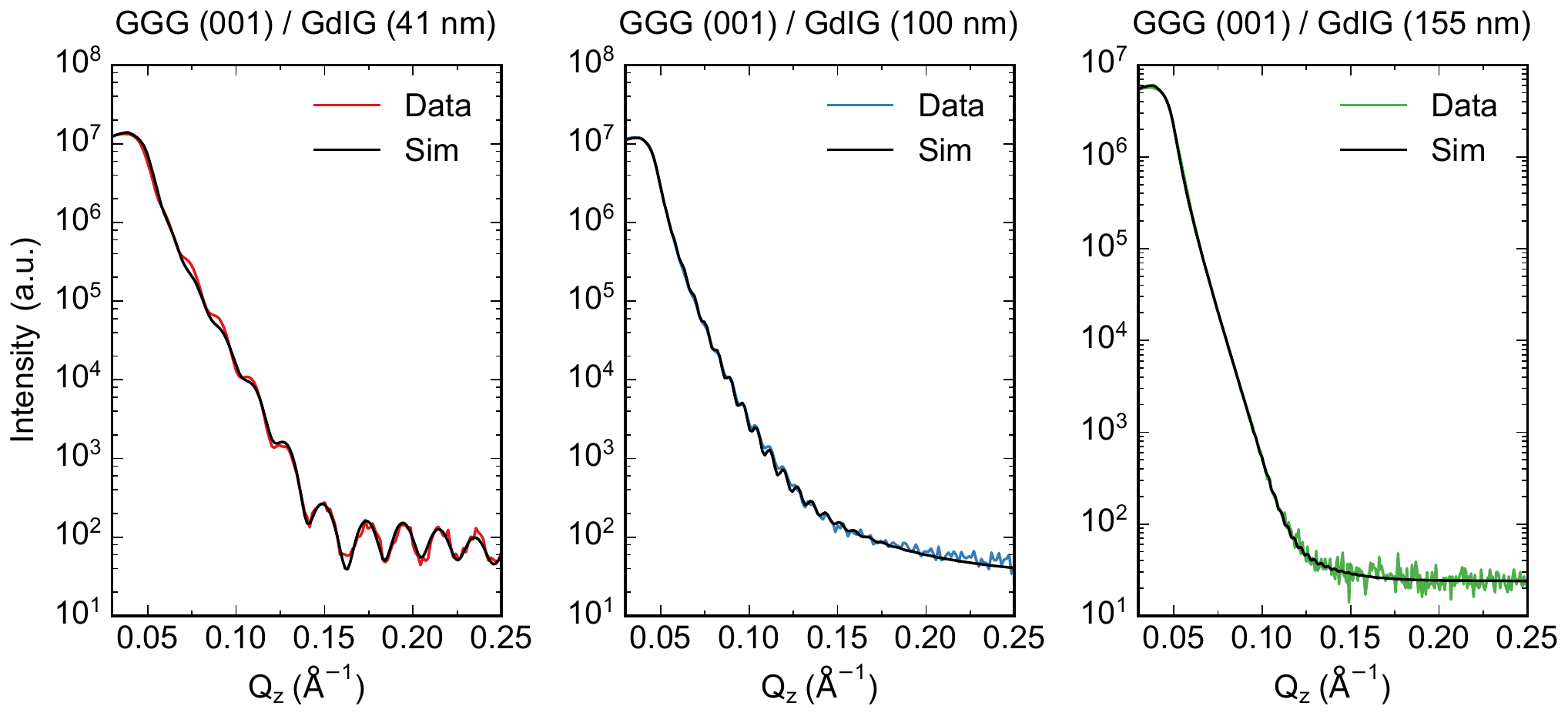}
	\caption{
		X-Ray reflectometry data of Gd\textsubscript{3}Fe\textsubscript{5}O\textsubscript{12} (GdIG) thin films of various thickness (\textit{d} = \SIlist{41;100;155}{\nano\meter}) investigated at JGU Mainz.}
	\label{fig:xrr_erjia}
\end{figure}

To obtain information about the thickness and surface roughness of the grown garnet films X-ray reflectometry (XRR) as well as atomic force microscopy (AFM) measurements are performed.
From the XRR data (Fig. \ref{fig:xrr_stephan} and \ref{fig:xrr_erjia}) the film thickness as well as the root mean square roughness $\sigma_{\mathrm{rms}}$ of the sample surface are extracted by fitting a simulation curve to the raw data.
In case of large thickness ()GGG/GdIG(\SI{155}{\nano\meter}, Fig. \ref{fig:xrr_erjia} (c)) no Kiessig fringes can be observed so that the film thickness is estimated by the growth time.

\begin{figure}[tb]
	\centering
	\includegraphics[width=0.4\textwidth]{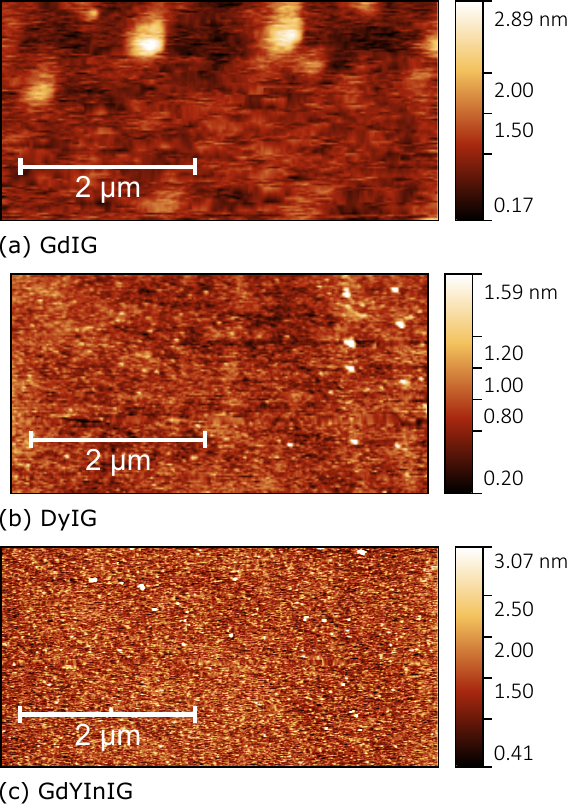}
	\caption{Atomic force microscopy data of rare earth iron garnets (a) Gd\textsubscript{3}Fe\textsubscript{5}O\textsubscript{12} (GdIG), (b) Dy\textsubscript{3}Fe\textsubscript{5}O\textsubscript{12} (DyIG), and (c) (Gd\textsubscript{2}Y\textsubscript{1})(Fe\textsubscript{4}In\textsubscript{1})O\textsubscript{12} (GdYInIG).}
	\label{fig:afm_gdig}
\end{figure}

The AFM measurements (Fig. \ref{fig:afm_gdig}) allow to extract $\sigma_{\mathrm{rms}}$ directly from the topographical information yielded by the data.
As seen in Fig. \ref{fig:afm_gdig} (a)-(c), the sample surfaces are partially or significantly contanimated by dirt particles resulting from sample handling, leading to an increase of $\sigma_{\mathrm{rms}}$ as compared to the values obtained by XRR.

The $\sigma_{\mathrm{rms}}$ values obtained by the different methods for the investigated films are summarized in Tab. \ref{tab:roughness}.

\begin{table}[bt]
	\centering
	\caption{Thickness and roughness data for the investigated garnet films obtained by XRR and AFM.}
	\vspace{0.25cm}
	\begin{tabular}{cccc}
		\toprule
		Film & $d$ $\left( \si{\nano\meter} \right)$ & $\sigma_{\mathrm{rms,XRR}}$ $\left( \si{\nano\meter} \right)$ & $\sigma_{\mathrm{rms,AFM}}$ $\left( \si{\nano\meter} \right)$\\
		\colrule
		YAG (111) / GdIG & 26 & 0.60 & 0.54\\
		YAG (111) / DyIG & 30 & 0.40 & 0.80\\
		YAG (111) / GdYInIG & 61 & 0.35 & 0.43\\
		GGG (001) / GdIG & 41 & 1.37 & -\\
		GGG (001) / GdIG & 100 & 1.71 & -\\
		GGG (001) / GdIG & 155 & 2.27 & -\\
		\botrule
	\end{tabular}
	\label{tab:roughness}
\end{table}

\clearpage
\FloatBarrier	
\subsection{Transmission electron microscopy characterization}

A direct insight into the crystalline structure of part of the investigated samples is obtained via high resolution scanning transmission microscopy (HR-STEM).
For HR-STEM the samples have been prepared using the standard focused ion beam (FIB) protocol.
Respective HR-STEM images and electron energy loss spectroscopy (EELS) spectra have been taken using an aberration-corrected scanning transmission electron microscope (FEI Probe Corrected Titan3\textsuperscript{TM} 80-300 S/TEM) operating at 200 kV and equipped with a Gatan Quantum ERS spectrometer.

\begin{figure}[bt]
	\centering
	\includegraphics[width=0.95\textwidth]{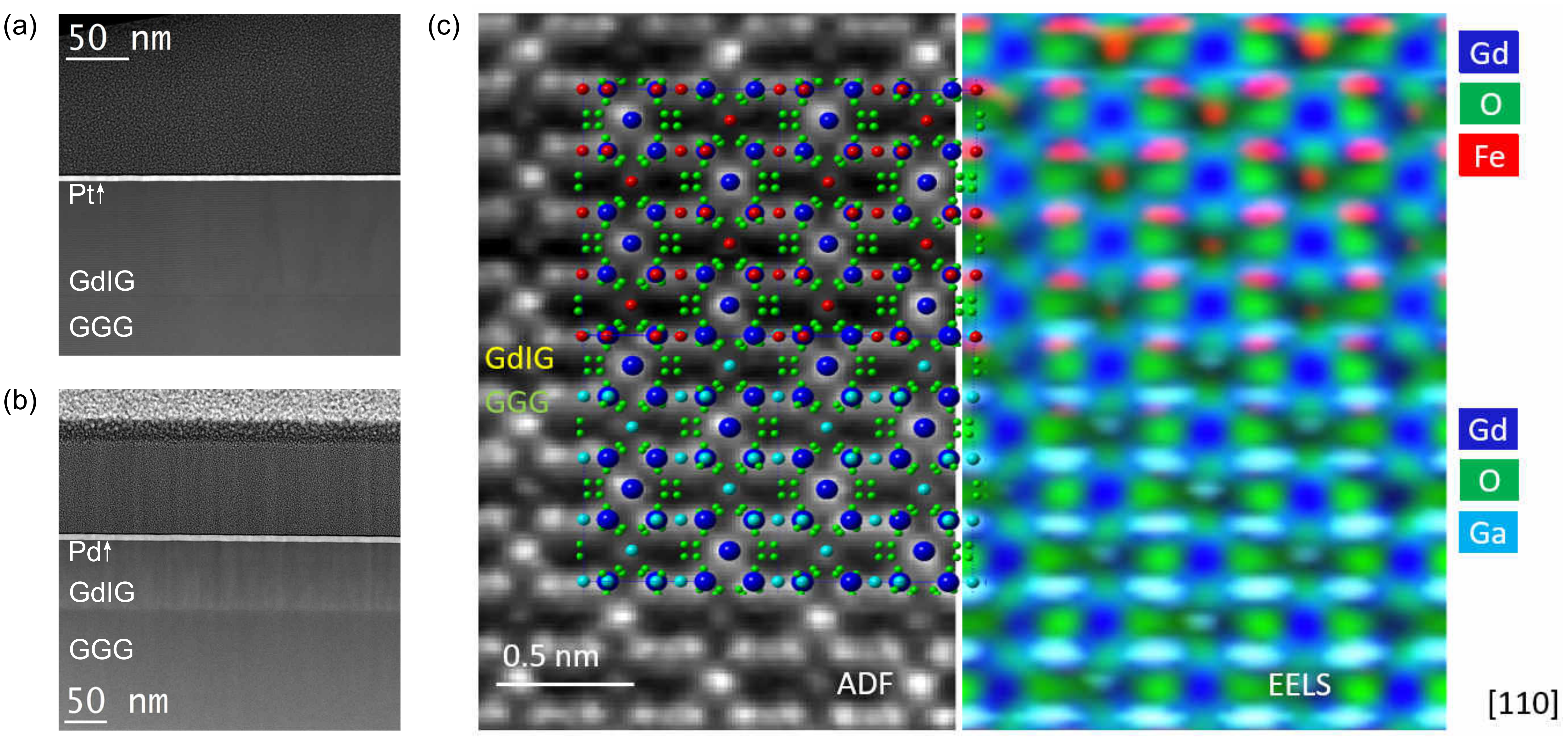}
	\caption{(a),(b) Annular dark field (ADF) images of (a) GGG(001)/GdIG/Pt and (b) GGG(001)/GdIG/Pd samples investigated at JGU Mainz.
		(c) ADF (left) and electron energy loss spectroscopy (EELS, right) image of the GGG/GdIG interface of the sample shown in (b).
		The images are taken alongside the [110] crystal orientation using a scanning transmission electron microscope.
		For comparison, the expected crystal structure is shown schematically showing good agreement. }
	\label{fig:tem_nice}
\end{figure}

Fig. \ref{fig:tem_nice} (a),(b) show two examples of high-angle annular dark field (HAADF) cross sections of the GGG(001)/GdIG/Pt and GGG(001)/GdIG/Pd samples, whose spin Seebeck data are presented in Fig. 3(a),(b) in the main text.
The images already indicate the epitaxial growth of the garnet films as well as reveal the overall flatness of the metal films.
In Fig. \ref{fig:tem_nice} (c) we show an image example of the interface between substrate and grown garnet film of the GGG(001)/GdIG/Pd, combining the information obtained by annular dark field (ADF) and electron energy loss spectroscopy (EELS) imaging.
The clear crystalline structure of substrate and grown film are pictured with atomic resolution, revealing the high epitaxial quality of the latter and a well defined interface.
The elemental maps given by the EELS data allow us to assign or rather identify elements at the lattice sites, which show little interdiffusion of Gd and Ga atoms between GGG and GdIG.

\begin{figure}[tb]
	\centering
	\includegraphics[width=0.7\textwidth]{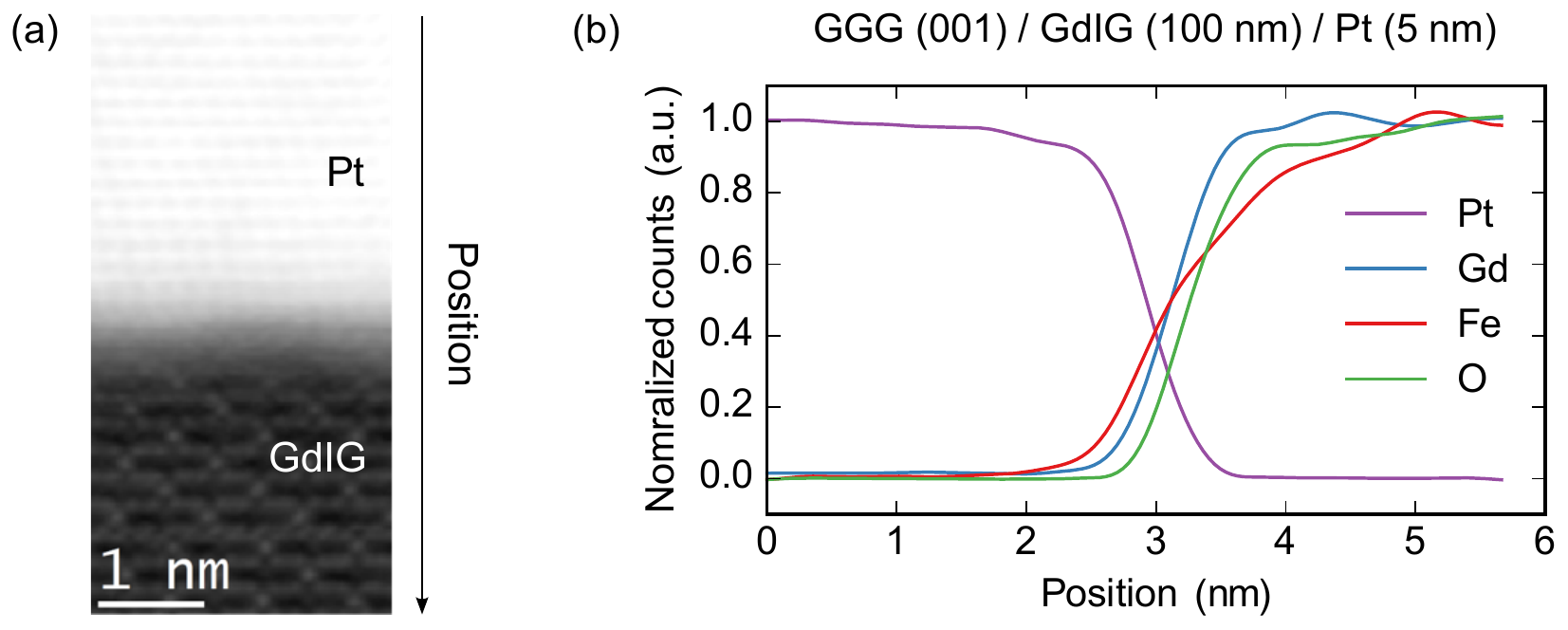}
	\caption{(a) High resolution scanning electron microscopy ADF image of the GdIG/Pt interface of the GGG(001)/GdIG/Pt sample investigated at JGU Mainz.
		Respective spin Seebeck data are shown in Fig. 3 in the main text.
		(b) Distribution of single elements across the interface obtained by an electron energy loss spectroscopy elemental map.
		The direction of the scan is indicated by the arrow in (a).}
	\label{fig:tem_PtPd}
\end{figure}

In Fig. \ref{fig:tem_PtPd} we show an example of a detailed HR-STEM images of the GdIG/Pt interface, including ADF information as well as information about the distribution of single elements via EELS.
On the one hand, the images reveal that the garnet film retains its crystalline structure up to at least \SI{100}{\nano\meter} thickness.
On the other hand, a sub-nanometer, Gd/Fe rich interface between Pt and GdIG is observed via the EELS elemental map.
In case of GdIG/Pd (not shown here) a Gd/O rich interface is observed.

\clearpage
\section{SQUID magnetometry of GdIG and DyIG}

\begin{figure}[h]
	\centering
	\includegraphics[width = 1\textwidth]{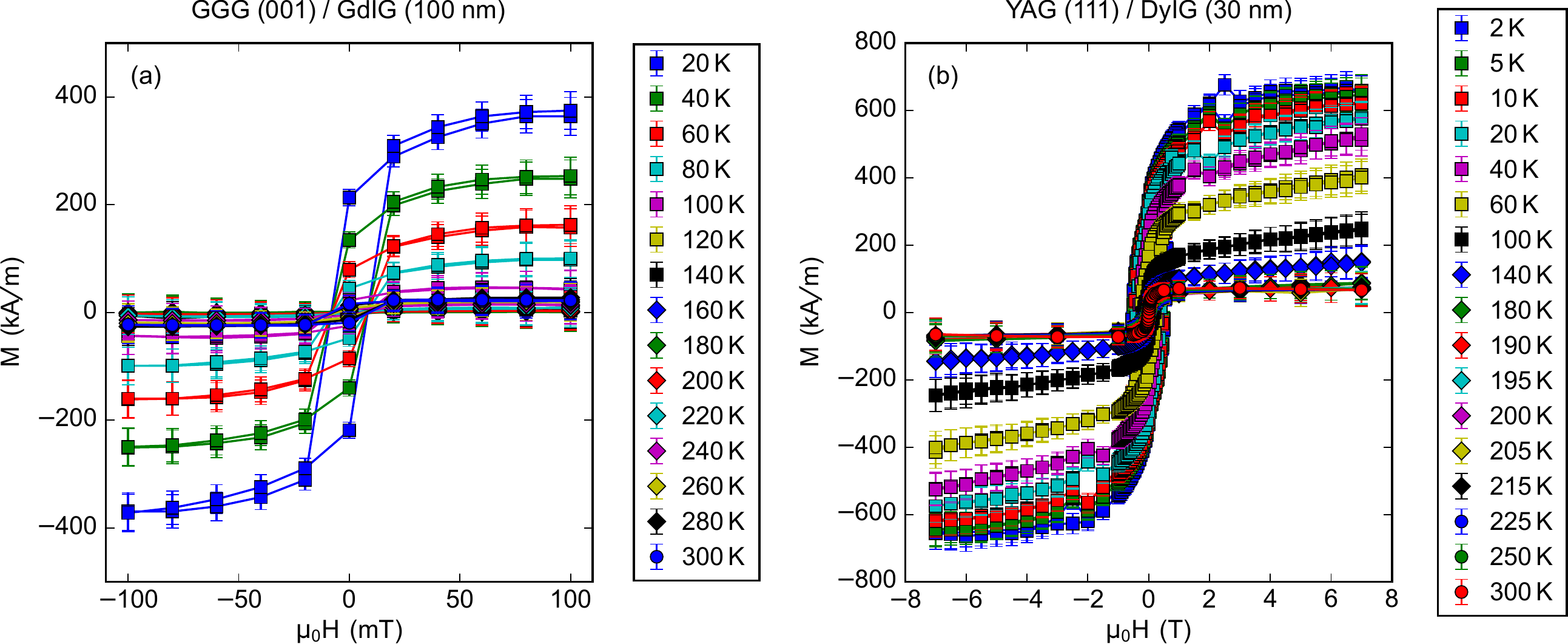}
	\caption{Magnetization of (a) GdIG and (b) DyIG as a function of external magnetic field at different temperatures.}
	\label{fig:m_vs_h}
\end{figure}

In Fig. \ref{fig:m_vs_h}, SQUID magnetometry data obtained for GGG(001)/GdIG(\SI{100}{\nano\meter}) and YAG(111)/DyIG(\SI{30}{\nano\meter}) are shown, depicting the sample magnetization as a function of external magnetic field at different temperatures.
While measuring the respective data, in addition to the magnetic moment of the grown iron garnets, the paramagnetic (diamagnetic) moment of GGG (YAG) is recorded as well.
To remove this contribution a linear fit is applied to the raw magnetiation data in the field range where the sample is assumed to be fully magnetized.
Subsequently the raw data are rectified by means of the obtained linear slope.
In case of YAG/DyIG the diamagnetic background is determined once for \SI{2}{\kelvin} and thereafter, assuming a constant diamagnetic contribution of YAG, all temperature datasets are corrected using the determined parameters.
For GGG/GdIG the paramagnetic background of GGG is calculated for each temperature.
Since the latter is large as compared to the GdIG magnetization, this procedure may lead to an error in the slope of $M\,(H)$ in the saturated state.

\vspace{2cm}

\clearpage
\section{Typical SSE temperature dependence measurements}

For a comprehensive characterization of the spin Seebeck (SSE) temperature dependence different measurements can be performed, here shown exemplarily for the GGG(001)/GdIG(\SI{100}{\nano\meter})/Pt(\SI{5}{\nano\meter}) sample measured at JGU Mainz.

\begin{figure}[h]
	\centering
	\includegraphics[width = 0.9\textwidth]{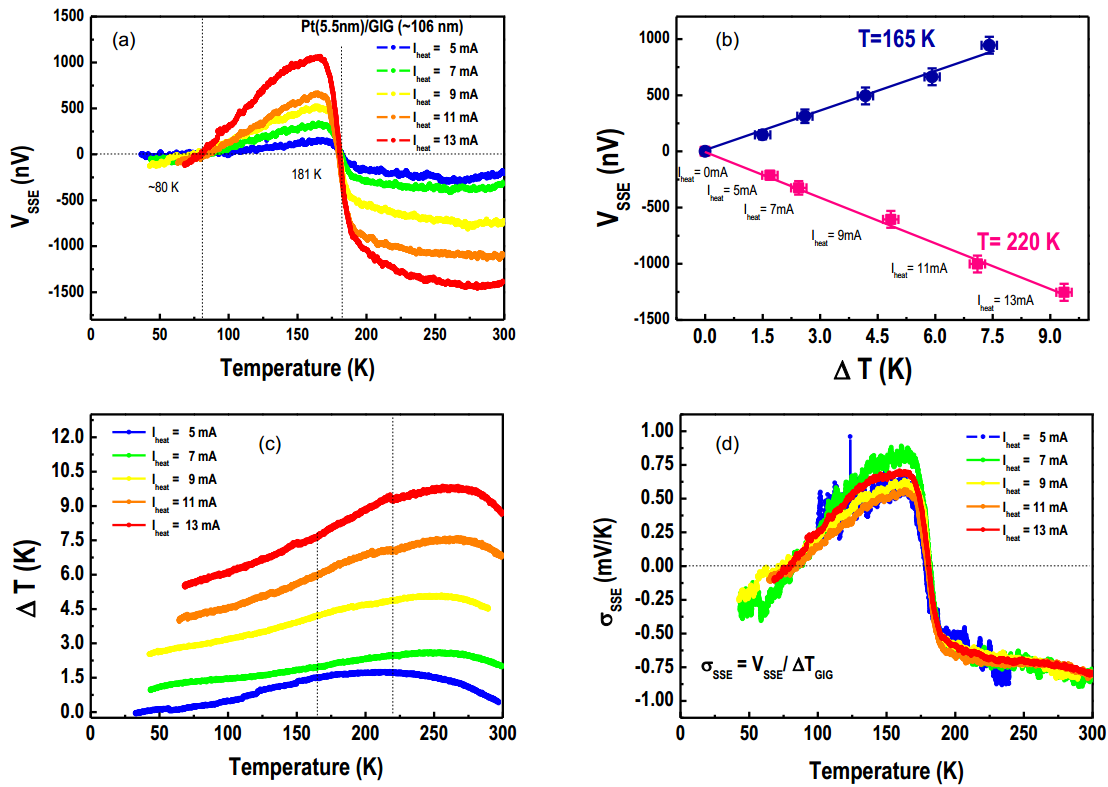}
	\caption{Typical data obtained from the GGG/GdIG (\SI{100}{\nano\meter}) sample during measurements using the setup at JGU Mainz.
		(a) Spin Seebeck voltage $V_{\mathrm{LSSE}}$ as a function of temperature for different applied heating currents.
		(b) $V_{\mathrm{LSSE}}$ as a function of temperature difference between sample top and bottom at different temperatures, revealing a linear dependence.
		(c) Temperature difference between sample top and bottom as a function of temperature for different applied heating currents.
		(d) Spin Seebeck coefficient $\sigma_{\mathrm{LSSE}} = V_{\mathrm{LSSE}}/ \Delta T$ as a function of temperature.}
	\label{fig:erjia}
\end{figure}

Fig. \ref{fig:erjia} (a) shows the temperature dependence of the recorded SSE voltage $V_{\mathrm{LSSE}}$ for different heating currents.
As expected, $V_{\mathrm{LSSE}}$ increases with increasing heating current.
Using the temperature calibration method described in the main text one can extract the dependence of $V_{\mathrm{LSSE}}$ on the temperature difference $\Delta T$ applied across the sample, revealing a linear dependence (Fig. \ref{fig:erjia} (b)).
This corresponds to the behavior expected for the SSE and is observed for both regimes above and below the magnetization compensation point where $V_{\mathrm{LSSE}}$ changes its sign.
Generally, the position of the first and second sign change of $V_{\mathrm{LSSE}}$ is independent on the applied heating power, emphasizing their physical significance and their dependence on intrinsic material properties.

Using the evaluated temperature difference $\Delta T$ across the sample (plotted in Fig. \ref{fig:erjia} (c), again for the different applied heating currents) one can deduce the temperature dependence of the spin Seebeck coefficient $\sigma_{\mathrm{LSSE}}$, see Fig. \ref{fig:erjia} (d).
Within the error the different curves obtained for different heating currents superimpose, demonstrating the robustness of the measurement method.

\vspace{2cm}

\clearpage

\end{document}